\documentclass{article}

\usepackage{arxiv}

\usepackage[utf8]{inputenc}
\usepackage[T1]{fontenc}
\usepackage{hyperref}
\usepackage{url}
\usepackage{booktabs}
\usepackage{amsfonts}
\usepackage{nicefrac}
\usepackage{microtype}
\usepackage{graphicx}
\usepackage{natbib}
\usepackage{doi}
\usepackage{amsmath,amssymb}
\usepackage{cleveref}
\usepackage{multirow}
\usepackage{subcaption}
\usepackage{xcolor}
\usepackage{pgfplots}
\pgfplotsset{compat=1.18}
\usepackage{threeparttable}
\usepackage{rotating}
\usepackage{float}
\usepackage{tabularx}
\newcolumntype{L}{>{\raggedright\arraybackslash}X}
\usepackage{placeins} 

\renewcommand{\headeright}{Preprint}
\renewcommand{\undertitle}{Preprint}
\renewcommand{\shorttitle}{Which CS1 Students Will Fail? Identifying Digital Markers from Learning Analytics}

\hypersetup{
  pdftitle={Which CS1 Students Will Fail? Identifying Digital Markers from Learning Analytics in Computer Systems and Architecture Using Weighted Academic Momentum and Interaction Logs},
  pdfsubject={Learning Analytics, Educational Data Mining},
  pdfauthor={Lighton Phiri, Mutune Chaibela, Ivy Chisha, David Pungwa, Danny Siabbaba, Bydon Simukoko},
  pdfkeywords={CS1, student success, predictive analytics, learning analytics, digital markers, early warning system, Moodle}
}

\title{Which CS1 Students Will Fail? Identifying Digital Markers from Learning Analytics in Computer Systems and Architecture Using Weighted Academic Momentum and Interaction Logs}

\author{
  \href{https://orcid.org/0000-0003-3582-9866}{\includegraphics[scale=0.06]{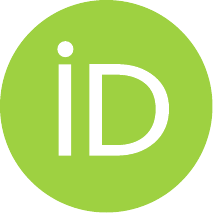}\hspace{1mm}Lighton Phiri}\thanks{Corresponding author: \texttt{lighton.phiri@unza.ac.zm}} \\
  Department of Computing and Informatics, University of Zambia, P.O Box 32379, Lusaka, Zambia \\
  DataLab Research Group, University of Zambia, P.O Box 32379, Lusaka, Zambia \\
  \And
  \href{}{\includegraphics[scale=0.06]{orcid.pdf}\hspace{1mm}Mutune Chaibela} \\
  Department of Computing and Informatics, University of Zambia, P.O Box 32379, Lusaka, Zambia \\
  DataLab Research Group, University of Zambia, P.O Box 32379, Lusaka, Zambia \\
  \And
  \href{}{\includegraphics[scale=0.06]{orcid.pdf}\hspace{1mm}Ivy Chisha} \\
  Department of Computing and Informatics, University of Zambia, P.O Box 32379, Lusaka, Zambia \\
  DataLab Research Group, University of Zambia, P.O Box 32379, Lusaka, Zambia \\
  \And
  \href{}{\includegraphics[scale=0.06]{orcid.pdf}\hspace{1mm}David Pungwa} \\
  Department of Computing and Informatics, University of Zambia, P.O Box 32379, Lusaka, Zambia \\
  DataLab Research Group, University of Zambia, P.O Box 32379, Lusaka, Zambia \\
  \And
  \href{}{\includegraphics[scale=0.06]{orcid.pdf}\hspace{1mm}Danny Siabbaba} \\
  Department of Computing and Informatics, University of Zambia, P.O Box 32379, Lusaka, Zambia \\
  DataLab Research Group, University of Zambia, P.O Box 32379, Lusaka, Zambia \\
  \And
  \href{}{\includegraphics[scale=0.06]{orcid.pdf}\hspace{1mm}Bydon Simukoko} \\
  Department of Computing and Informatics, University of Zambia, P.O Box 32379, Lusaka, Zambia \\
  DataLab Research Group, University of Zambia, P.O Box 32379, Lusaka, Zambia \\
}

\begin{document}

\maketitle

\begin{abstract}
Digital learning platforms generate rich behavioural traces – digital markers – that offer the potential to identify struggling students early. This paper investigates whether a combination of traditional and digital markers can predict failure in a first‑year CS1 course (Computer Systems and Architecture) with sufficient recall to enable timely intervention. Using data from four cohorts (2017–2021, N=284) at a large public university in sub‑Saharan Africa, we conducted a mixed‑methods stakeholder elicitation (surveys, interviews, focus groups) to identify ten candidate factors. These were operationalised into a comprehensive feature set spanning demographics, self‑reported surveys, Moodle interaction logs, and continuous assessment scores. A systematic ablation study using logistic regression with 5‑fold cross‑validation and SMOTE+ENN resampling revealed that the most predictive feature subset was `Base + Demo + LMS`: weighted academic momentum ($M = 0.1Q_1 + 0.15Q_2 + 0.2Q_3 + 0.55T_1$), basic demographics (gender, sponsorship, COVID‑19 cohort), and a binary indicator of any LMS activity. On a held‑out test set, logistic regression achieved 74.7\% accuracy, 0.742 macro F1, and an AUC of 0.800. At the default threshold of 0.5, the model identified 87\% of failing students (recall = 0.87) with a 41\% false positive rate – a trade‑off suitable for early intervention. SHAP analysis confirmed that weighted academic momentum is the strongest predictor, followed by its interaction with LMS engagement; self‑sponsored status and high course workload emerged as risk factors. These results demonstrate that simple digital markers – early assessment scores and a binary LMS engagement flag – can be systematically identified and validated to power a practical early‑warning system that detects the vast majority of at‑risk CS1 students by the fifth week of the semester. The main contributions are: (1) a multi‑source dataset and a stakeholder‑guided methodology for identifying digital markers; (2) an ablation study that quantifies the contribution of different feature groups; and (3) an interpretable, high‑recall model ready for operational deployment.
\end{abstract}


\section{Introduction}
\label{sec:introduction}

The rapid digitalisation of higher education has produced an unprecedented volume of behavioural data – often called \emph{digital markers} – that can be leveraged to improve student outcomes. Learning management systems (LMS) such as Moodle record every click, every forum post, and every minute spent on course materials. When combined with traditional data sources (demographics, prior grades, assessment scores), these digital markers offer a powerful lens through which instructors can identify struggling students early enough to intervene.

This potential is particularly acute in introductory computer science courses (CS1), where failure rates are notoriously high. At the authors' institution, the CS1 course \emph{Computer Systems and Architecture} has recorded an average failure rate of 39.7\% over four cohorts. Despite decades of pedagogical innovations, many students still struggle, and instructors often lack timely information about who is at risk.

The central thesis of this paper is that digital markers, when appropriately engineered into predictive models, can enable early detection of at‑risk CS1 students with sufficient recall to support practical intervention. To investigate this, we address five research questions. First, to what extent can a combination of digital markers – demographics, self‑reported survey responses, LMS interaction logs, and continuous assessment scores – predict failure in a CS1 course (RQ1)? Second, which digital markers are the strongest individual predictors, and how do they interact, for example between momentum and LMS engagement (RQ2)? Third, what is the maximum recall for the failing class achievable by a logistic regression model after threshold tuning, and what is the corresponding false positive rate (RQ3)? Fourth, how does the inclusion of digital markers (LMS logs and self‑reported factors) improve performance over using only formative assessment scores (RQ4)? Fifth, what actionable insights can instructors derive from SHAP analysis to inform early interventions (RQ5)?

In addressing these questions, the main contributions of this paper are:
\begin{itemize}
    \item A rich, multi‑source dataset (anonymised and made available to researchers) that captures four cohorts of CS1 students.
    \item A novel \emph{weighted academic momentum} score that weights early quizzes and the first test according to their instructional importance.
    \item An empirical comparison of seven classifiers, with systematic handling of class imbalance (SMOTE+ENN) and threshold tuning.
    \item An interpretable logistic regression model that achieves 72\% accuracy and 87\% recall for the failing class, accompanied by SHAP analysis to explain predictions.
    \item A discussion of how this model can be operationalised as an early‑warning system in real CS1 classrooms.
\end{itemize}

The remainder of the paper is structured as follows. Section~\ref{sec:related_work} reviews the landscape of predictive analytics in CS1. Section~\ref{sec:methodology} provides a detailed account of the data collection, feature engineering, and modelling pipeline. Section~\ref{sec:results} presents descriptive statistics, model performance, and the visualisation and interpretation of digital markers via clustering, heatmaps, ROC curves, and SHAP. Section~\ref{sec:discussion} interprets the findings in the context of digital marker research and outlines practical intervention strategies. Section~\ref{sec:conclusion} concludes and suggests future directions.

\section{Related Work}
\label{sec:related_work}

The landscape of Artificial Intelligence in Education (AIED) has evolved rapidly over the last decade, transitioning from descriptive analysis to advanced predictive modelling aimed at improving learning outcomes \cite{Guo2024, Ersozlu2024}. Predicting student performance remains a cornerstone of educational data mining (EDM), with research shifting toward the use of large‑scale datasets and complex algorithmic implementations to identify students at risk of failure or dropout \cite{Guo2024, Batool2023}. A meta‑analysis of over 260 studies indicates that while Artificial Neural Networks (ANN) and Random Forests (RF) are prevalent, the efficacy of these models is heavily dependent on the quality of input features and the handling of dataset specificities \cite{Batool2023}.

\subsection{Digital Markers and Feature Engineering in CS1}
Early identification of at‑risk students is particularly critical in high‑failure courses such as CS1. Recent studies have highlighted the limitations of relying solely on static demographic data, which often prove to be insufficient indicators of a student's underlying learning obstacles \cite{Shafiq2022, Hellas2018Predicting}. Instead, the inclusion of digital markers—behavioural traces captured by Learning Management Systems (LMS)—has been shown to significantly enhance predictive power \cite{Shafiq2022}. \citet{Alhazmi2023} demonstrated that combining admission scores with first‑level course grades yields more accurate early predictions than student characteristics alone. Furthermore, \citet{Hasan2020Predicting} identified that temporal engagement patterns, such as video lecture interaction, are strong predictors of risk. Despite these advances, there remains a gap in the literature regarding the creation of theoretically motivated composite scores, such as weighted academic momentum, which integrate early formative assessment trends with behavioural engagement \cite{Hellas2018Predicting}.

\subsection{Algorithmic Performance and Class Imbalance}
While various supervised learning algorithms have been evaluated, Support Vector Machines (SVM) and Decision Trees (DT) consistently show high performance when properly tuned. \citet{Ahmed2024} reported reaching 96\% accuracy using SVM with hyperparameter optimization, though \citet{Batool2023} noted that irrelevant features can significantly degrade performance and increase processing time. A critical but often overlooked challenge is class imbalance, where passing students vastly outnumber those failing. \citet{Shafiq2022} observed that only 21\% of surveyed studies addressed this issue, leading to models with high overall accuracy but poor recall for the at‑risk minority class. This study addresses this gap by utilizing SMOTE+ENN to balance classes and employing threshold tuning to maximize recall, a strategy proven essential for actionable early‑warning systems \cite{Susnjak2024, Ahmed2024}.

\subsection{From Interpretability to Prescriptive Analytics}
Modern EDM research increasingly argues that "black‑box" models are insufficient for educational contexts; they must be interpretable for instructors to act upon them \cite{Susnjak2024, Ersozlu2024}. SHAP (Shapley additive explanations) \cite{NIPS2017_7062} has emerged as a standard for realizing both global and local transparency, allowing researchers to quantify the influence of specific digital markers on individual predictions \cite{Susnjak2024, Hellas2018Predicting}. Furthermore, the field is moving toward prescriptive analytics, which goes beyond identifying \emph{what} will happen to suggest \emph{how} to change the outcome \cite{Susnjak2024}. Recent frameworks have integrated counterfactual modelling and large language models (LLMs) like ChatGPT to generate evidence‑based, remedial advice for at‑risk learners \cite{Susnjak2024}. By applying SHAP selection and threshold tuning, our work builds on this trend to provide instructors with both high‑recall alerts and actionable insights.

\section{Methodology}
\label{sec:methodology}

This study employed an exploratory sequential mixed‑methods design \cite{Creswell2017ResearchDesign}. This design is particularly suited to developing predictive models in educational contexts because it begins with qualitative exploration to identify relevant factors (here, barriers to student success), followed by quantitative measurement and modelling of those factors. The time horizon was longitudinal, spanning four academic cohorts (2017/18, 2018/19, 2019/20, 2020/21) to capture stable patterns as well as the disruption caused by the COVID‑19 pandemic. The implementation followed the CRISP‑DM (CRoss Industry Standard Process for Data Mining) framework \cite{Wirth1995CRISPDM}, with an additional front‑end stakeholder elicitation phase that directly informed the business understanding.

\subsection{Study Context and Participants}

The study was conducted at a large public university in sub‑Saharan Africa. The target course, CS1 S+A (Computer Systems and Architecture), is a mandatory first‑year course for all students in the Bachelor of Information and Communication Technologies with Education. It spans two semesters and covers computer systems organisation, architecture, and introductory programming concepts. Assessment is split 50/50 between continuous assessment (20 weekly quizzes, 4 semester tests) and a final examination. Table~\ref{tab:assessment_distribution} summarises the assessment weights.

\begin{table}[!htbp]
\caption{CS1 Assessment Distribution.}\label{tab:assessment_distribution}
\centering
\small
\begin{tabular}{lccc}
\toprule
Assessment & \% of Course & Frequency & Quantity \\
\midrule
Quiz  & 20\% & Weekly & 20 \\
Test  & 30\% & Semester & 4 \\
Examination & 50\% & Annual & 1 \\
\bottomrule
\end{tabular}
\end{table}

Data were collected for four cohorts: 2017/18, 2018/19, 2019/20, and 2020/21. The 2019/20 and 2020/21 cohorts experienced COVID‑19 disruptions (switch to fully online teaching); we include a binary flag to control for this.

\subsection{Stakeholder Elicitation of Factors}
\label{sec:stakeholder_factors}

To ground the predictive models in the pedagogical realities of the course rather than relying solely on available data, we conducted a three‑step stakeholder elicitation process that served as the foundational “Business/Domain Understanding” step of CRISP‑DM.

\subsubsection{Qualitative exploration (interviews and focus groups)}
First, we conducted semi‑structured interviews with the course instructor and two CS1 tutors, as well as two focus groups with senior students who had previously taken the course (total 12 participants). Thematic analysis of the transcripts and field notes identified ten distinct factors that participants believed influenced student performance: Time Management, Teaching Mode, Programme Workload, Prior Knowledge, Motivation, Lack of Orientation, Lack of Computer, Lack of Equipment, Interest, and Assessment Structure.

\subsubsection{Quantitative survey}
Second, to assess the prevalence of these factors across the broader student population, we developed a structured survey instrument. The survey was administered to a convenience sample of current and past CS1 students (N=29). Participants rated the impact of each factor on their academic performance using a 5‑point Likert scale (Strongly Disagree to Strongly Agree). Descriptive statistical analysis ranked the factors by severity. “Time Management”, “Motivation”, and “Lack of Computer” received the highest combined “Agree/Strongly Agree” ratings (each >75\%), while “Assessment Structure” was the only factor rated as not influential by a majority (>50\% Disagree/Strongly Disagree). These quantitative results corroborated the qualitative findings and provided a severity ranking (see Figure~\ref{fig:student_factors} in the Results section).

\subsubsection{Mapping to feature set}
Third, the validated factors were mapped directly onto our data mining pipeline. The high impact of “Lack of Computer” and “Motivation” justified the inclusion of pre‑course survey variables such as `SurveyOwnComputer`, `SurveyMajorMotivation`, and `SurveyMinorMotivation`. The importance of “Programme Workload” led to the inclusion of `CourseWorkload` from the Student Information System. “Prior Knowledge” was operationalised through the pre‑course survey items on computing experience and prior training. This mapping ensured that the final feature set was not a “black box” of convenient variables but instead was intrinsically linked to the lived experiences of students and instructors, thereby enhancing the interpretability and actionability of the resulting early‑warning system.

\subsection{Dataset Preparation}

\subsubsection{Data sources}
Four primary sources were used to construct the analytical dataset:
\begin{enumerate}
    \item \textbf{Student Information System (SIS)}: Provided demographic data (age, gender, sponsorship, accommodation, programme, minor, course workload).
    \item \textbf{Pre‑course survey}: Administered online at the beginning of each academic year. Collected self‑reported information on prior computing experience, computer ownership, motivation, and perceived barriers.
    \item \textbf{Moodle Learning Management System (LMS)}: Raw event logs were aggregated to daily unique hits per student. The log file contained 17,870 records with fields `StudentID`, `AcademicYear`, `MoodleDate`, `Component`, `Event name`, etc.
    \item \textbf{Assessment spreadsheets}: Provided individual scores for 20 weekly quizzes, four semester tests, and the final examination.
\end{enumerate}

\subsubsection{Data pre‑processing}
The following steps were performed before merging:

\paragraph{Renaming and cleaning columns.}
Column names were standardised to lower camel case (e.g., `StudentID`, `AcademicYear`). Survey response columns were renamed from question text to concise identifiers (e.g., `SurveyStudyComputersHighschool`). No duplicate columns were present after renaming.

\paragraph{Removal of duplicate records.}
No duplicate student records (by `StudentID` and `AcademicYear`) were found in the SIS or survey data. Moodle logs contained multiple rows per student per day (one per event); we aggregated these to daily unique hits, thereby removing event‑level duplicates.

\paragraph{Merging of multiple sources.}
The final dataset was created by left joins on `StudentID` and `AcademicYear` in the following order:
\begin{enumerate}
    \item Base: assessment spreadsheets (scores).
    \item Join SIS demographic data.
    \item Join pre‑course survey responses (using `StudentID` only, as surveys were not cohort‑specific).
    \item Join Moodle engagement metrics (total unique days, component counts) computed from the log file.
\end{enumerate}
After merging, fewer than 5\% of records had missing values in some columns due to incomplete surveys or missing LMS logs.

\paragraph{Handling missing values.}
Numeric features were imputed with the median of the respective feature; categorical features were imputed with the mode. Textual survey responses were used only for qualitative analysis and not as predictive features.

\paragraph{Data type conversions.}
`DateOfBirth` was converted to age at the start of the academic year. `AcademicYear` was stored as a string but converted to an ordered categorical for plotting; for modelling only a binary COVID‑19 flag was used. Likert scale responses were treated as ordered categorical and converted to numeric codes (1–5) for polynomial features.

\paragraph{Outlier detection.}
No outliers were removed because extreme values (e.g., very low quiz scores, very high Moodle hits) were considered meaningful signals. The maximum unique days (221) was plausible for a full academic year.

\subsubsection{Dataset description}
After pre‑processing, the clean dataset contained 284 complete student records (after excluding those with missing final exam scores). Table~\ref{tab:full_dataset} presents the complete set of features, organised by factor group (DEMO, COSP, COAC, COAS). For each feature, we indicate its source and its Data Mining Attribute Type – original, text, categorical, dichotomous, or numeric (ratio or interval). This typology follows standard data mining practice and clarifies how each feature was handled.

\begin{table*}[!htbp]
\centering
\caption{CS1 Course Dataset Description with Data Mining Attribute Types.}
\label{tab:full_dataset}
\small
\begin{tabularx}{\textwidth}{l L L L}
\toprule
\textbf{Factor} & \textbf{Feature} & \textbf{Source} & \textbf{Data Mining Attribute Type} \\
\midrule
\multirow{15}{*}{DEMO} & DateOfBirth & Student Information System & Numeric (ratio) – age derived \\
& Gender & Student Information System & Categorical (nominal) \\
& MajorDescription & Student Information System & Categorical \\
& MinorDescription & Student Information System & Categorical \\
& Sponsor & Student Information System & Categorical (sponsorship type) \\
& CampusAccommodation & Student Information System & Dichotomous \\
& SurveyHomeTownSuburb & Self‑administered Survey & Text (free response) \\
& SurveyProgramMinor & Self‑administered Survey & Text \\
& SurveyMinorMotivation & Self‑administered Survey & Text \\
& SurveyMajorMotivation & Self‑administered Survey & Text \\
& SurveyStudyComputersHighschool & Self‑administered Survey & Dichotomous (Yes/No) \\
& SurveyPriorComputerTraining & Self‑administered Survey & Dichotomous \\
& SurveyPriorComputerTrainingDetails & Self‑administered Survey & Text \\
& SurveyExperienceUsingComputers & Self‑administered Survey & Categorical (ordered: <1y,1-2y,3-5y,>5y) \\
& SurveyOwnComputer & Self‑administered Survey & Dichotomous \\
\hline
\multirow{4}{*}{COSP} & CoursesTaken & Student Information System & Text (course codes) \\
& MinorProgram & Student Information System & Categorical \\
& MinorClassification & Student Information System & Categorical (STEM/Social Science) \\
& CourseWorkload & Student Information System & Numeric (ratio) – total credits \\
\hline
\multirow{12}{*}{COAC} & MoodleHits & Learning Management System & Numeric (ratio) – unique days active \\
& MoodleHitsWeight & Learning Management System & Numeric (ratio) – proportion of total possible days \\
& MoodleLogComponentAssignment & Learning Management System & Numeric (ratio) – count \\
& MoodleLogComponentChoice & Learning Management System & Numeric (ratio) – count \\
& MoodleLogComponentFile & Learning Management System & Numeric (ratio) – count \\
& MoodleLogComponentFolder & Learning Management System & Numeric (ratio) – count \\
& MoodleLogComponentForum & Learning Management System & Numeric (ratio) – count \\
& \texttt{MoodleLogComponent\-OverviewReport} & Learning Management System & Numeric (ratio) – count \\
& MoodleLogComponentSystem & Learning Management System & Numeric (ratio) – count \\
& MoodleLogComponentURL & Learning Management System & Numeric (ratio) – count \\
& MoodleLogComponentUserReport & Learning Management System & Numeric (ratio) – count \\
& MoodleLogComponentUserTours & Learning Management System & Numeric (ratio) – count \\
\hline
\multirow{3}{*}{COAS} & Quiz N Score (N=1..20) & Spreadsheets & Numeric (interval) – percentage scores \\
& Test N Score (N=1..4) & Spreadsheets & Numeric (interval) – percentage scores \\
& Final Exam Score & Spreadsheets & Numeric (interval) – percentage score \\
\bottomrule
\end{tabularx}
\end{table*}

\subsection{Feature Engineering}

\subsubsection{Target Variable}
A student was labelled as failing (0) if the final examination score was below 45\%, otherwise passing (1). This cut‑off corresponds to the university’s minimum passing grade.

\subsubsection{Weighted Academic Momentum}
Early assessments in the CS1 course are designed to be cumulative: each subsequent quiz builds on previous topics, and the first test covers the material from the first three quizzes plus additional content. Therefore, a simple average of early scores would underweight the more informative later assessments. To address this, we constructed a weighted momentum score that gives progressively higher weight to later assessments:
\begin{equation}
M = 0.1 Q_1 + 0.15 Q_2 + 0.2 Q_3 + 0.55 T_1,
\end{equation}
where $Q_i$ are the first three quiz scores (each 0–100) and $T_1$ is the first test score (also 0–100). The specific weights (0.1, 0.15, 0.2, 0.55) were chosen through iterative consultation with the course instructor, based on the following pedagogical rationale: later quizzes cover more advanced material and are temporally closer to the first test, thus they should receive higher weight. The first test ($T_1$) is the most comprehensive early assessment – it synthesises several weeks of material – and therefore receives the dominant weight (0.55). The weights sum to 1, which maintains interpretability as a weighted average of the early assessments.

\subsubsection{Deriving Digital Markers from Moodle Logs}
The raw Moodle logs contained fields such as `Time`, `User full name`, `Affected user`, `Event context`, `Component`, `Event name`, `Description`, `Origin`, and `IP address`. A sample of raw log entries is shown in Table~\ref{tab:sample_logs_full}, using a reduced font size and wrapped columns to accommodate all fields.

\begin{table}[!htbp]
\centering
\caption{Sample raw Moodle log entries (anonymised, all fields).}
\label{tab:sample_logs_full}
\tiny
\setlength{\tabcolsep}{2pt}
\begin{tabularx}{\textwidth}{llllllXll}
\toprule
\textbf{Time} & \textbf{User full name} & \textbf{Affected user} & \textbf{Event context} & \textbf{Component} & \textbf{Event name} & \textbf{Description} & \textbf{Origin} & \textbf{IP address} \\
\midrule
28/05/21, 08:42 & [Instructor] & - & Course: ICT 1110 Comp. Sys. \& Arch. (2020/21) & Logs & Log report viewed & The user with id '14785' viewed the log report for the course with id '2413'. & web & 165.56.182.238 \\
28/05/21, 08:41 & [Instructor] & - & Course: ICT 1110 Comp. Sys. \& Arch. (2019/20) & System & Course viewed & The user with id '14785' viewed the course with id '2413'. & web & 165.56.182.238 \\
\bottomrule
\end{tabularx}
\end{table}

To transform these raw event logs into usable digital markers, we performed the following aggregations:

\begin{itemize}
    \item \textbf{Binary engagement indicator (`has\_lms\_record`)}: For each student, we recorded whether there was at least one log entry over the entire semester. Students without any recorded interaction (no rows in the log file) received a value of 0. This binary flag captures the most basic form of digital presence.
    \item \textbf{Total unique days active (`MoodleHits`)}: From the `Time` field, we extracted the date and counted the number of distinct days on which a student generated at least one log event. This metric reflects the frequency of engagement over time.
    \item \textbf{Component‑specific counts}: By parsing the `Component` field (e.g., “Quiz”, “Forum”, “File”, “URL”), we counted how many times a student accessed each type of resource. These counts were used to explore fine‑grained behavioural patterns.
\end{itemize}

After aggregation, the derived digital markers were joined with the main dataset using `StudentID` and `AcademicYear`. From these markers, we further engineered higher‑level features to capture non‑linear and interactive effects:

\begin{itemize}
    \item \textbf{Polynomial features}: Degree‑2 polynomial terms were added for `MoodleHits`, `CourseWorkload`, and the weighted academic momentum $M$. This allows the model to detect curvilinear relationships (e.g., a U‑shaped effect of LMS engagement on failure risk).
    \item \textbf{Interaction term}: We explicitly created the product $\texttt{M\_times\_LMS} = M \times \texttt{has\_lms\_record}$. This interaction captures the conditional effect of early academic momentum in the presence or absence of any LMS activity – a key insight identified during stakeholder interviews.
\end{itemize}

\subsubsection{Encoding of Categorical Variables}
Nominal categorical features (`Gender`, `MinorDescription`) were one‑hot encoded using \texttt{pandas.get\_dummies()} with \texttt{drop\_first=True} to avoid multicollinearity. Ordinal categorical features (e.g., `SurveyExperienceUsingComputers`) were left as numeric codes.

\subsubsection{Standardization}
All numeric features were standardised (zero mean, unit variance) using the \texttt{StandardScaler} from scikit-learn before being passed to classifiers that assume normally distributed inputs (logistic regression, SVC, ANN). Tree‑based models (Random Forest, XGBoost, LightGBM) used the raw unscaled values.

After imputation, one‑hot encoding, polynomial expansion, and standardisation, the final feature matrix contained 47 columns. This is substantially different from the original 69 raw columns because we first removed non‑predictive columns: student identifiers (`StudentID`, `StudentName`), free‑text survey fields used only in the qualitative stakeholder analysis, and the target variable `FinalExamination` (which is not a feature). After this removal, we had 34 raw predictive attributes (those listed in Table~\ref{tab:full_dataset}). These 34 raw attributes were then transformed: one‑hot encoding of categorical variables produced multiple dummy columns; degree‑2 polynomial features for `MoodleHits`, `CourseWorkload`, and $M$ added derived features; and the interaction term contributed one extra column. The net result was an increase from 34 raw predictors to 47 engineered features ready for modelling.

\subsection{Class Imbalance and Resampling}
The dataset had 186 passing (65.5\%) and 98 failing (34.5\%) students. To prevent models from simply predicting the majority class, we applied SMOTE+ENN (Synthetic Minority Over‑sampling + Edited Nearest Neighbours) inside a 5‑fold stratified cross‑validation loop. SMOTE creates synthetic examples of the minority class; ENN removes noisy samples from both classes. This combination reduces overfitting compared to SMOTE alone.

\subsection{Model Training and Hyperparameter Tuning}
We evaluated seven classifiers: Logistic Regression (LR), Random Forest (RF), XGBoost (XGB), LightGBM (LGB), SVC (RBF kernel), a stacking ensemble (LR + RF + XGB as base learners, LR as meta‑learner), and a shallow Artificial Neural Network (ANN) with two hidden layers (64 and 32 neurons, ReLU activation, dropout 0.4, Adam optimiser). Hyperparameters were tuned using \texttt{RandomizedSearchCV} with 3‑fold cross‑validation. The search spaces are shown in Table~\ref{tab:hyperparams}.

\begin{table}[!htbp]
\caption{Hyperparameter search spaces.}\label{tab:hyperparams}
\centering
\small
\begin{tabular}{ll}
\toprule
Model & Hyperparameters \\
\midrule
Random Forest & \texttt{n\_estimators} $\in \{100,200,300\}$, \texttt{max\_depth} $\in \{5,10,15,\text{None}\}$, \texttt{min\_samples\_split} $\in \{2,5,10\}$ \\
XGBoost & \texttt{n\_estimators} $\in \{100,200,300\}$, \texttt{max\_depth} $\in \{3,5,7\}$, \texttt{learning\_rate} $\in \{0.01,0.05,0.1\}$ \\
\bottomrule
\end{tabular}
\end{table}

All models were trained on the training fold after applying SMOTE+ENN, then evaluated on the original test fold. Performance metrics were averaged over the 5 folds.

\subsection{Threshold Tuning}
Because our primary goal is to identify at‑risk students (i.e., maximise recall for the failing class), we tuned the decision threshold. For each model, we varied the threshold from 0.1 to 0.9 in steps of 0.05 and selected the threshold that maximised macro F1 on the validation set. For logistic regression, the optimal threshold was 0.65.

\subsection{Model Interpretability with SHAP}
To understand which features drive predictions and to provide actionable insights for instructors, we used SHAP (SHapley Additive exPlanations) with the XGBoost model as a surrogate. SHAP values decompose each prediction into additive feature contributions, satisfying local accuracy, missingness, and consistency. We generated summary plots and dependence plots to identify interactions.

\subsection{Experimental Evaluation and Ablation Study}

\subsubsection{Evaluation Metrics}
We assessed model performance using standard binary classification metrics:
\begin{itemize}
    \item \textbf{Accuracy}: proportion of correctly classified instances (both pass and fail).
    \item \textbf{Macro F1 score}: harmonic mean of precision and recall, averaged equally across both classes. Robust to class imbalance.
    \item \textbf{AUC (Area Under the ROC Curve)}: model’s ability to discriminate between passing and failing students, independent of a threshold.
    \item \textbf{Recall (Sensitivity) for the failing class}: proportion of actual failing students correctly identified. This is our primary optimisation target.
    \item \textbf{Precision for the failing class}: proportion of students flagged as failing who actually fail.
    \item \textbf{False Positive Rate (FPR) for the passing class}: proportion of passing students incorrectly flagged as failing.
\end{itemize}
All reported results are from a 30\% hold‑out test set after performing 5‑fold stratified cross‑validation on the training set.

\subsubsection{Ablation Study: Validating Stakeholder Factors}
The stakeholder elicitation (Section~\ref{sec:stakeholder_factors}) identified several factors as potentially influential: time management, motivation, prior computing experience, lack of computer, programme workload, and course engagement. To quantitatively verify the contribution of each factor group, we conducted an ablation study using logistic regression (the best‑performing classifier in our preliminary experiments). We trained the model on nested subsets of features, starting from a minimal baseline and progressively adding feature groups that correspond to stakeholder claims.

The feature groups are defined as follows:

\begin{itemize}
    \item \textbf{Base}: Only the weighted academic momentum score ($M$). This represents the purely formative‑assessment baseline.
    \item \textbf{+Demo+LMS}: Adds demographic features (gender, sponsorship, COVID‑19 cohort flag) and LMS activity markers (has\_lms\_record, MoodleHits). This tests the impact of engagement and basic demographics.
    \item \textbf{+CourseReg}: Adds course registration features (course workload, minor classification, minor programme). This tests the stakeholder claim that programme workload affects performance.
    \item \textbf{+SurvCat}: Adds survey categorical features (computer ownership, prior computer training, high school computing experience, self‑rated computing experience level). This tests the influence of prior exposure and access to equipment.
    \item \textbf{+SurvText}: Adds TF‑IDF vectors extracted from the free‑text motivation responses (“SurveyMinorMotivation” and “SurveyMajorMotivation”). This tests whether students’ expressed motivation contains predictive signal beyond structured answers.
    \item \textbf{+Interact}: Adds all pairwise interaction terms between the previous groups (e.g., gender $\times$ workload, momentum $\times$ has\_lms\_record, sponsorship $\times$ computer ownership). This captures non‑additive effects that stakeholders might not articulate directly.
\end{itemize}

The ablation uses the same train/test split (70/30) and SMOTE+ENN resampling applied only to the training folds. For each feature set, we train a logistic regression model and evaluate accuracy, macro F1, and AUC on the test set. The results, presented in Section~\ref{sec:ablation_results} (in the Results section), directly indicate whether adding a group of stakeholder‑aligned features improves predictive performance beyond the formative baseline.

\section{Results}
\label{sec:results}

\subsection{Descriptive Analysis of Student Performance}
\label{sec:descriptive}

\begin{figure}[!htbp]
\centering
\includegraphics[width=0.8\textwidth]{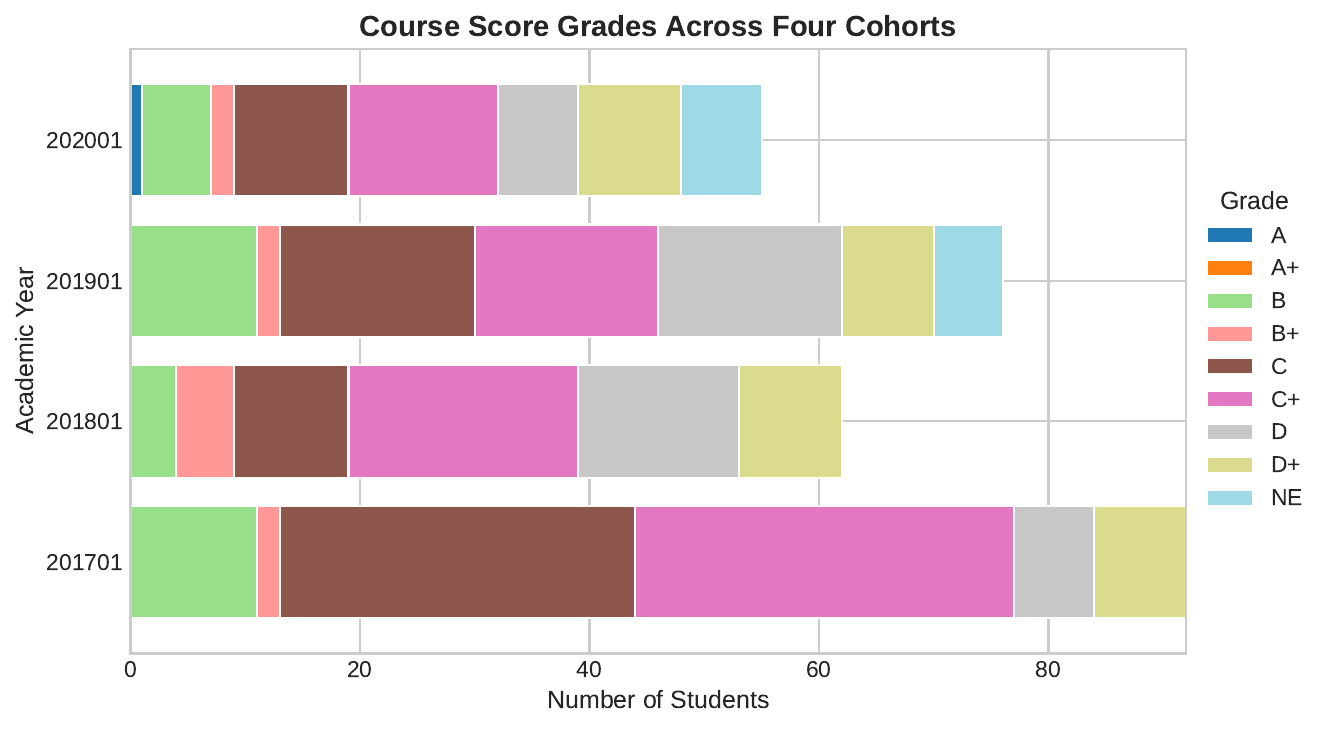}
\caption{Course score grades across four cohorts.}
\label{fig:course_grades}
\end{figure}

Figure~\ref{fig:course_grades} shows the distribution of final course grades across the four cohorts. The 2019/20 cohort (COVID‑19) shows a marked increase in NE (Not Examined) and D grades. The overall failure rate (D+, D, NE) averaged 39.7\%, confirming the severity of the problem.

\begin{figure}[!htbp]
\centering
\includegraphics[width=0.9\textwidth]{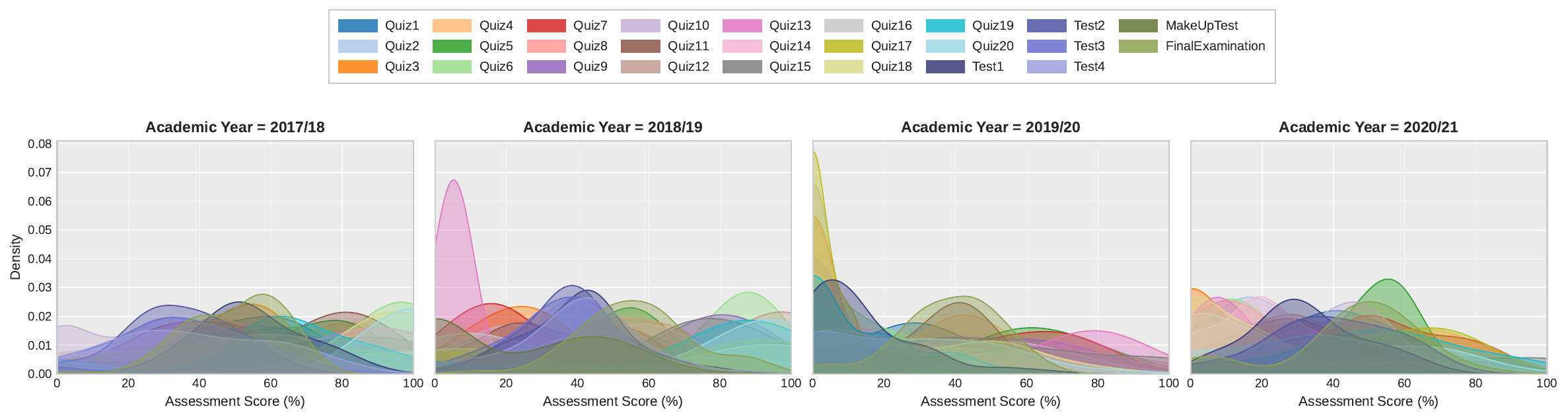}
\caption{Distribution of assessment scores across cohorts.}
\label{fig:assessment_distributions}
\end{figure}

Figure~\ref{fig:assessment_distributions} presents kernel density estimates for all 20 quizzes, 4 tests, and the final examination. Several patterns emerge: quizzes 6, 12, and 18 show extreme bimodality (many students score very high or very low). The COVID‑19 cohorts (2019/20, 2020/21) exhibit a shift toward lower scores, especially in quizzes 6–12, which coincided with the lockdown period. Test 1 scores are generally higher than later tests, reflecting the introductory nature of early material. The final examination distribution is right‑skewed, with a long left tail of failures.

\FloatBarrier
\begin{table*}[!htbp]
\centering
\caption{Descriptive statistics of assessment scores by cohort (failure rates, mean, median, min, max).}
\label{tab:descriptive_stats}
\scriptsize
\setlength{\tabcolsep}{2.0pt}
\resizebox{\textwidth}{!}{
\begin{tabular}{l *{24}{c}}
\toprule
 & \multicolumn{6}{c}{2017/18} & \multicolumn{6}{c}{2018/19} & \multicolumn{6}{c}{2019/20} & \multicolumn{6}{c}{2020/21} \\
 \cmidrule(lr){2-7} \cmidrule(lr){8-13} \cmidrule(lr){14-19} \cmidrule(lr){20-25}
 & \rotatebox{90}{Pass (\%)}  & \rotatebox{90}{Fail (\%)} & $\bar{x}$ & $\tilde{x}$ & \rotatebox{90}{$\max(x)$}  & \rotatebox{90}{$\min(x)$} & \rotatebox{90}{Pass (\%)}  & \rotatebox{90}{Fail (\%)} & $\bar{x}$ & $\tilde{x}$ & \rotatebox{90}{$\max(x)$}  & \rotatebox{90}{$\min(x)$} & \rotatebox{90}{Pass (\%)}  & \rotatebox{90}{Fail (\%)} & $\bar{x}$ & $\tilde{x}$ & \rotatebox{90}{$\max(x)$}  & \rotatebox{90}{$\min(x)$} & \rotatebox{90}{Pass (\%)}  & \rotatebox{90}{Fail (\%)} & $\bar{x}$ & $\tilde{x}$ & \rotatebox{90}{$\max(x)$}  & \rotatebox{90}{$\min(x)$}   \\
\midrule
Exam & 64 & 36 & 49.9 & 52.5 & 74.0 & 5.0 & 80.0 & 20 & 56.5 & 55.8 & 90.0 & 12.0 & 37 & 63 & 38.6 & 40.0 & 65.0 & 2.0 & 61 & 39 & 45.1 & 48.5 & 80.8 & 1.0 \\ \midrule
Test1 & 74 & 26 & 52.5 & 52.0 & 88.0 & 0.0 & 36.0 & 64 & 40.1 & 42.0 & 80.0 & 0.0 & 4 & 96 & 13.3 & 9.6 & 65.9 & 0.0 & 23 & 77 & 32.5 & 30.0 & 66.0 & 4.0 \\ \midrule
Test2 & 74 & 26 & 34.4 & 34.0 & 70.0 & 0.0 & 36.0 & 64 & 36.6 & 38.0 & 64.0 & 0.0 & --- & --- & 0.0 & 0.0 & 0.0 & 0.0 & 23 & 77 & 39.7 & 40.0 & 73.0 & 0.0 \\ \midrule
Test3 & 24 & 76 & 35.1 & 34.0 & 84.0 & 0.0 & 24.0 & 76 & 35.3 & 36.0 & 68.0 & 0.0 & --- & --- & 0.0 & 0.0 & 0.0 & 0.0 & --- & --- & 0.0 & 0.0 & 0.0 & 0.0 \\ \midrule
Test4 & 40 & 60 & 38.7 & 38.0 & 88.0 & 0.0 & 38.0 & 62 & 40.0 & 42.0 & 86.0 & 0.0 & --- & --- & 0.0 & 0.0 & 0.0 & 0.0 & --- & --- & 0.0 & 0.0 & 0.0 & 0.0 \\ \midrule
Quiz1 & 68 & 32 & 53.1 & 60.0 & 90.0 & 0.0 & 46.0 & 54 & 37.0 & 40.0 & 100.0 & 0.0 & 19 & 81 & 25.1 & 25.0 & 70.0 & 0.0 & 35 & 65 & 38.1 & 40.0 & 60.0 & 0.0 \\ \midrule
Quiz2 & 68 & 32 & 47.6 & 50.0 & 100.0 & 0.0 & 46.0 & 54 & 35.7 & 40.0 & 90.0 & 0.0 & 19 & 81 & 8.1 & 0.0 & 70.0 & 0.0 & 35 & 65 & 22.0 & 20.0 & 90.0 & 0.0 \\ \midrule
Quiz3 & 66 & 34 & 48.6 & 50.0 & 80.0 & 0.0 & 14.0 & 86 & 27.1 & 27.5 & 70.0 & 0.0 & 4 & 96 & 5.5 & 0.0 & 70.0 & 0.0 & 7 & 93 & 11.5 & 0.0 & 80.0 & 0.0 \\ \midrule
Quiz4 & 76 & 24 & 62.1 & 80.0 & 90.0 & 0.0 & 46.0 & 54 & 36.8 & 40.0 & 75.0 & 0.0 & 28 & 72 & 30.7 & 37.5 & 62.5 & 0.0 & 67 & 33 & 52.8 & 57.5 & 94.0 & 0.0 \\ \midrule
Quiz5 & 91 & 9 & 66.6 & 70.0 & 100.0 & 0.0 & 79.0 & 21 & 57.1 & 55.0 & 100.0 & 0.0 & 70 & 30 & 48.9 & 56.7 & 93.0 & 0.0 & 79 & 21 & 52.1 & 54.5 & 85.0 & 12.0 \\ \midrule
Quiz6 & 88 & 12 & 84.0 & 100.0 & 100.0 & 0.0 & 92.0 & 8 & 80.0 & 90.0 & 100.0 & 0.0 & 18 & 82 & 15.4 & 10.0 & 50.0 & 0.0 & 16 & 84 & 20.0 & 10.0 & 80.0 & 0.0 \\ \midrule
Quiz7 & 46 & 54 & 39.3 & 40.0 & 85.0 & 0.0 & 11.0 & 89 & 22.3 & 20.0 & 80.0 & 0.0 & 61 & 39 & 45.9 & 53.3 & 80.0 & 0.0 & 77 & 23 & 54.3 & 52.3 & 85.7 & 0.0 \\ \midrule
Quiz8 & 46 & 54 & 40.6 & 40.0 & 90.0 & 0.0 & 63.0 & 38 & 48.0 & 50.0 & 80.0 & 0.0 & 25 & 75 & 30.0 & 20.0 & 100.0 & 0.0 & 14 & 86 & 18.7 & 10.0 & 80.0 & 0.0 \\ \midrule
Quiz9 & 63 & 37 & 49.7 & 50.0 & 90.0 & 0.0 & 86.0 & 14 & 70.7 & 80.0 & 100.0 & 0.0 & 45 & 55 & 42.0 & 39.3 & 91.3 & 0.0 & 42 & 58 & 37.9 & 38.3 & 80.0 & 0.0 \\ \midrule
Quiz10 & 11 & 89 & 23.6 & 30.0 & 90.0 & 0.0 & 29.0 & 71 & 32.5 & 30.0 & 80.0 & 0.0 & 40 & 60 & 34.3 & 30.0 & 90.0 & 0.0 & 47 & 53 & 37.8 & 44.5 & 70.0 & 0.0 \\ \midrule
Quiz11 & 89 & 11 & 71.9 & 80.0 & 100.0 & 0.0 & 38.0 & 63 & 33.8 & 30.0 & 60.0 & 0.0 & 18 & 82 & 29.5 & 35.5 & 56.7 & 0.0 & 26 & 74 & 33.0 & 30.8 & 80.8 & 0.0 \\ \midrule
Quiz12 & 75 & 25 & 64.2 & 70.0 & 100.0 & 0.0 & 88.0 & 13 & 83.2 & 100.0 & 100.0 & 0.0 & 40 & 60 & 40.4 & 40.0 & 100.0 & 0.0 & 7 & 93 & 21.4 & 24.5 & 50.0 & 0.0 \\ \midrule
Quiz13 & 80 & 20 & 70.9 & 80.0 & 100.0 & 0.0 & 2.0 & 98 & 7.1 & 5.0 & 60.0 & 0.0 & 63 & 37 & 51.9 & 80.0 & 100.0 & 0.0 & 5 & 95 & 19.2 & 10.0 & 50.0 & 0.0 \\ \midrule
Quiz14 & 55 & 45 & 44.8 & 50.0 & 100.0 & 0.0 & 64.0 & 36 & 52.0 & 60.0 & 90.0 & 0.0 & 25 & 75 & 24.0 & 15.0 & 90.0 & 0.0 & 5 & 95 & 20.4 & 19.5 & 50.0 & 0.0 \\ \midrule
Quiz15 & 72 & 28 & 65.1 & 80.0 & 100.0 & 0.0 & 86.0 & 14 & 67.9 & 70.0 & 100.0 & 0.0 & 58 & 42 & 45.1 & 54.5 & 100.0 & 0.0 & 53 & 47 & 49.9 & 45.0 & 100.0 & 0.0 \\ \midrule
Quiz16 & 64 & 36 & 52.3 & 60.0 & 100.0 & 0.0 & 36.0 & 64 & 33.9 & 17.5 & 100.0 & 0.0 & 3 & 97 & 4.0 & 0.0 & 50.0 & 0.0 & 19 & 81 & 18.6 & 10.0 & 70.0 & 0.0 \\ \midrule
Quiz17 & 60 & 40 & 52.7 & 50.0 & 100.0 & 0.0 & 57.0 & 43 & 57.3 & 90.0 & 100.0 & 0.0 & 1 & 99 & 3.9 & 0.0 & 50.0 & 0.0 & 65 & 35 & 52.3 & 54.7 & 87.1 & 0.0 \\ \midrule
Quiz18 & 84 & 16 & 78.8 & 90.0 & 100.0 & 0.0 & 52.0 & 48 & 52.5 & 90.0 & 100.0 & 0.0 & 34 & 66 & 27.9 & 20.0 & 80.0 & 0.0 & 9 & 91 & 13.1 & 0.0 & 60.0 & 0.0 \\ \midrule
Quiz19 & 85 & 15 & 60.3 & 60.0 & 100.0 & 0.0 & 84.0 & 16 & 73.0 & 90.0 & 100.0 & 0.0 & 3 & 97 & 11.6 & 0.0 & 60.0 & 0.0 & 63 & 37 & 52.1 & 52.3 & 100.0 & 0.0 \\ \midrule
Quiz20 & 84 & 16 & 84.1 & 100.0 & 100.0 & 0.0 & 80.0 & 20 & 76.8 & 90.0 & 100.0 & 0.0 & 27 & 73 & 26.4 & 25.0 & 90.0 & 0.0 & 33 & 67 & 37.9 & 40.0 & 90.0 & 0.0 \\
\bottomrule
\end{tabular}
}
\end{table*}
\FloatBarrier

\subsection{Stakeholder Factors}
\label{sec:stakeholder_factors_results}
\begin{figure}[!htbp]
\centering
\includegraphics[width=0.85\textwidth]{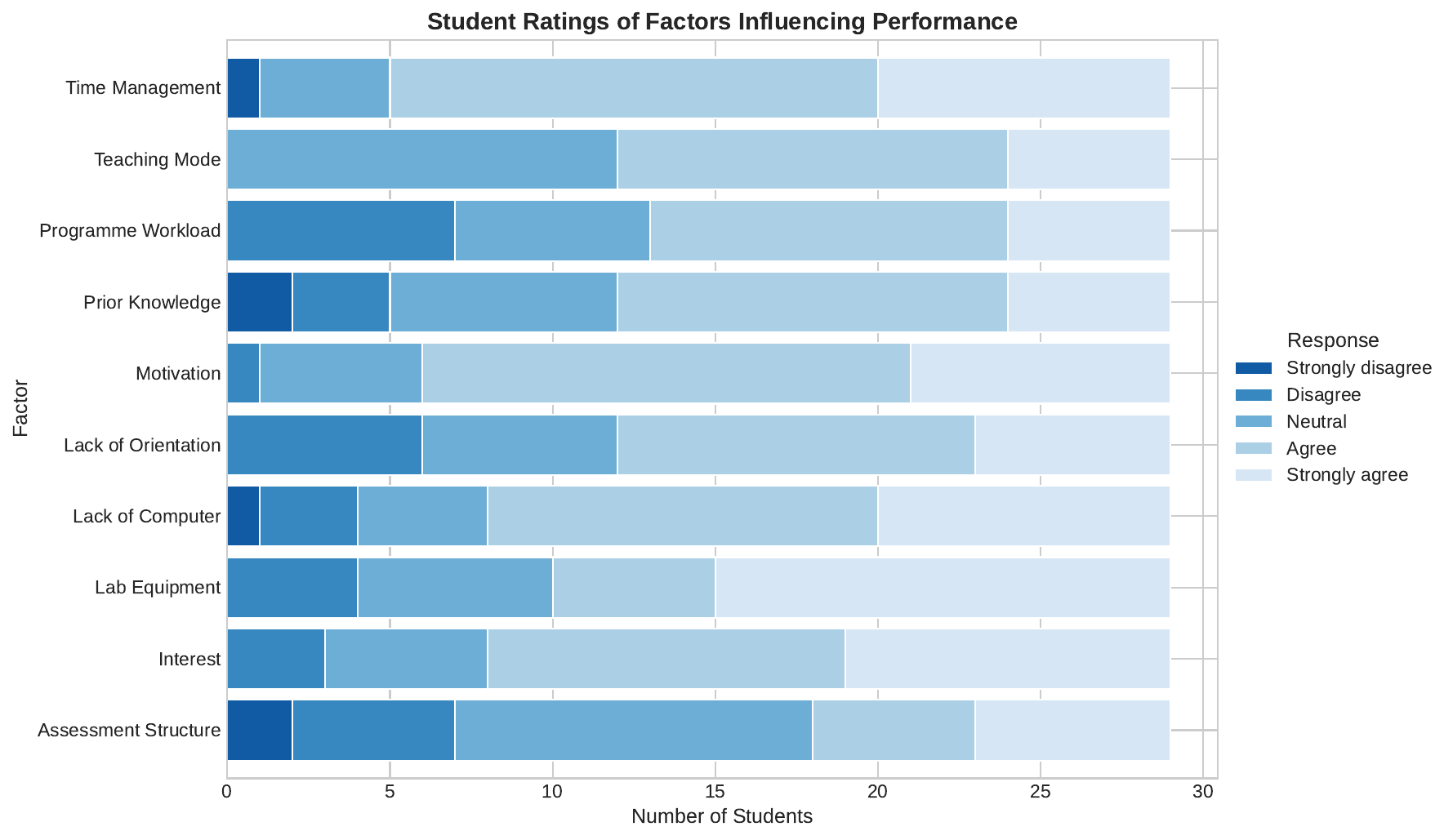}
\caption{Student ratings of factors influencing performance.}
\label{fig:student_factors}
\end{figure}

Figure~\ref{fig:student_factors} shows student ratings of ten factors identified through focus groups. “Time Management”, “Motivation”, and “Lack of Computer” received the highest combined “Agree/Strongly agree” ratings. “Assessment Structure” was the only factor rated as not influential by a majority. These results guided our inclusion of self‑reported computer ownership and sponsorship (as a proxy for financial resources) as features.

\subsection{Digital Marker Analysis: Moodle Engagement}
\label{sec:moodle_analysis}

To understand the relationship between LMS engagement and final performance, we performed two complementary analyses: K‑means clustering of total unique Moodle hits against final exam score, and a monthly heatmap of engagement bands.

\subsubsection{K‑Means Clustering}
\begin{figure}[!htbp]
\centering
\includegraphics[width=0.7\textwidth]{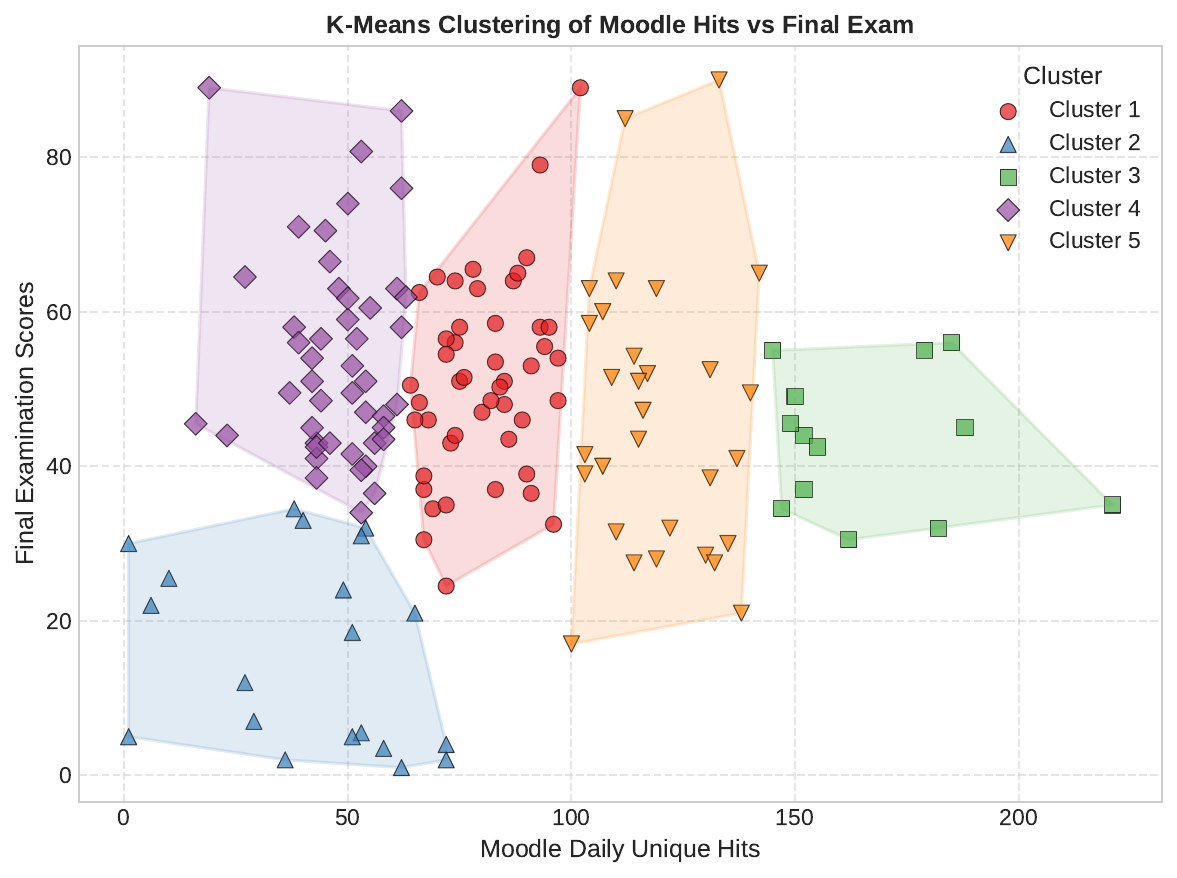}
\caption{K‑means clustering of Moodle daily unique hits vs. final examination score.}
\label{fig:moodle_clustering}
\end{figure}

Figure~\ref{fig:moodle_clustering} shows a K‑means clustering (k=5, silhouette score 0.32) of students based on total unique Moodle hits and final examination score. The clusters reveal a U‑shaped relationship: students in the lowest engagement cluster (0) predominantly fail, while those in moderate engagement clusters (1–3) show a wide range of performance. Notably, a small number of students with very high hits (cluster 4) also tend to fail, suggesting obsessive but unproductive engagement (e.g., repeatedly checking grades without studying course materials). This non‑linear relationship supports the inclusion of both a binary engagement flag and polynomial features in our model.

\subsubsection{Heatmap of Engagement Bands Over Time}
\begin{figure}[!htbp]
\centering
\includegraphics[width=0.8\textwidth]{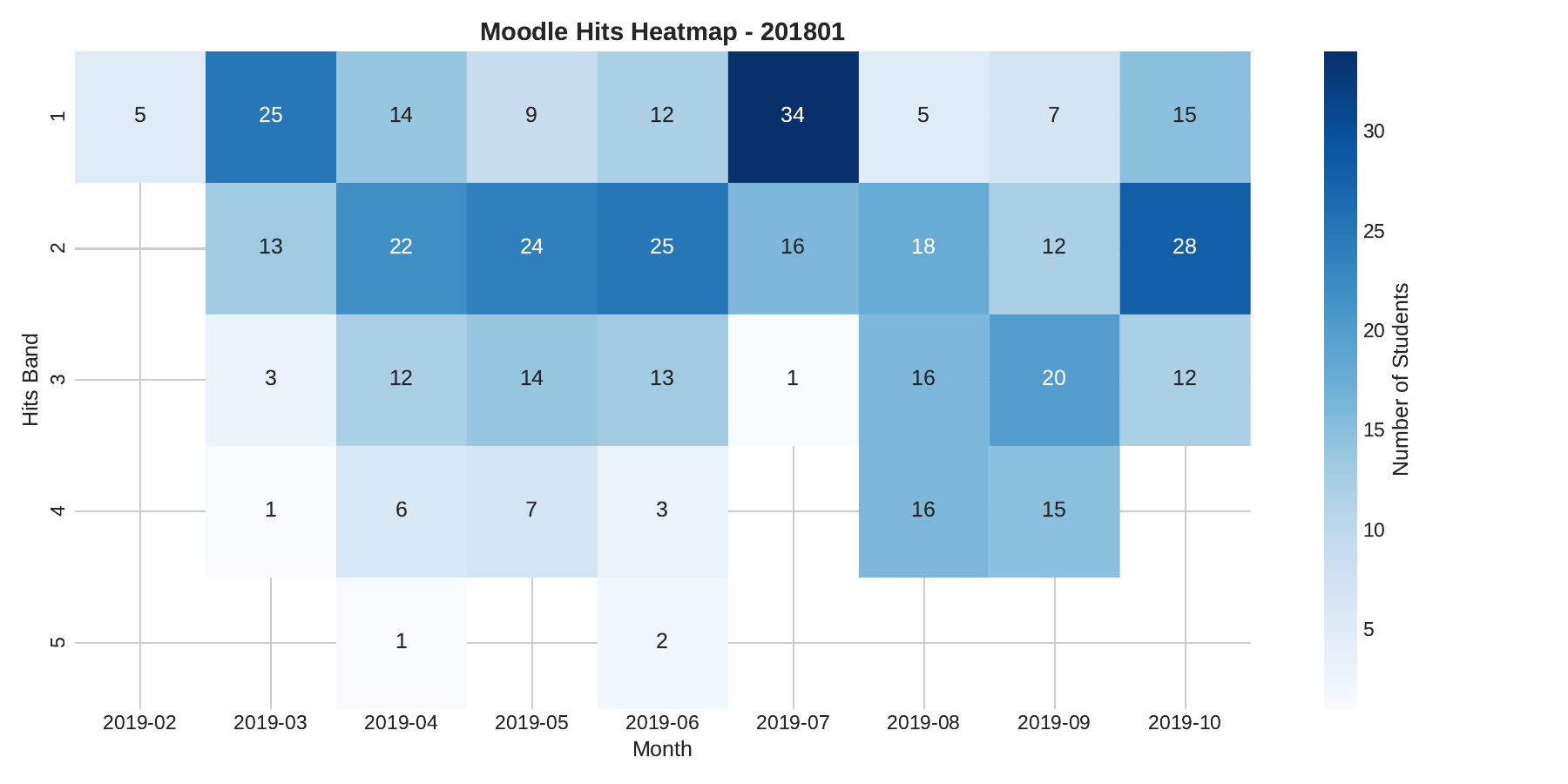}
\caption{Heatmap of Moodle hits bands per month for the 2018/19 cohort.}
\label{fig:moodle_heatmap}
\end{figure}

To capture temporal patterns, we divided the academic year into months and grouped students into seven “hits bands” (1: 1–5 hits/month, …, 7: >30). Figure~\ref{fig:moodle_heatmap} shows a heatmap for the 2018/19 cohort. Early months (September–October) show many students in high bands (5–7), but by November–December, a substantial proportion drop to bands 1–2. The COVID‑19 cohorts (not shown) displayed even sharper declines during lockdown months. This decline in engagement – which typically precedes the final examination by several weeks – is a strong digital marker of risk.

\subsection{Model Performance}
\label{sec:model_performance}

\subsubsection{Ablation Study}
\begin{figure}[!htbp]
\centering
\includegraphics[width=0.8\textwidth]{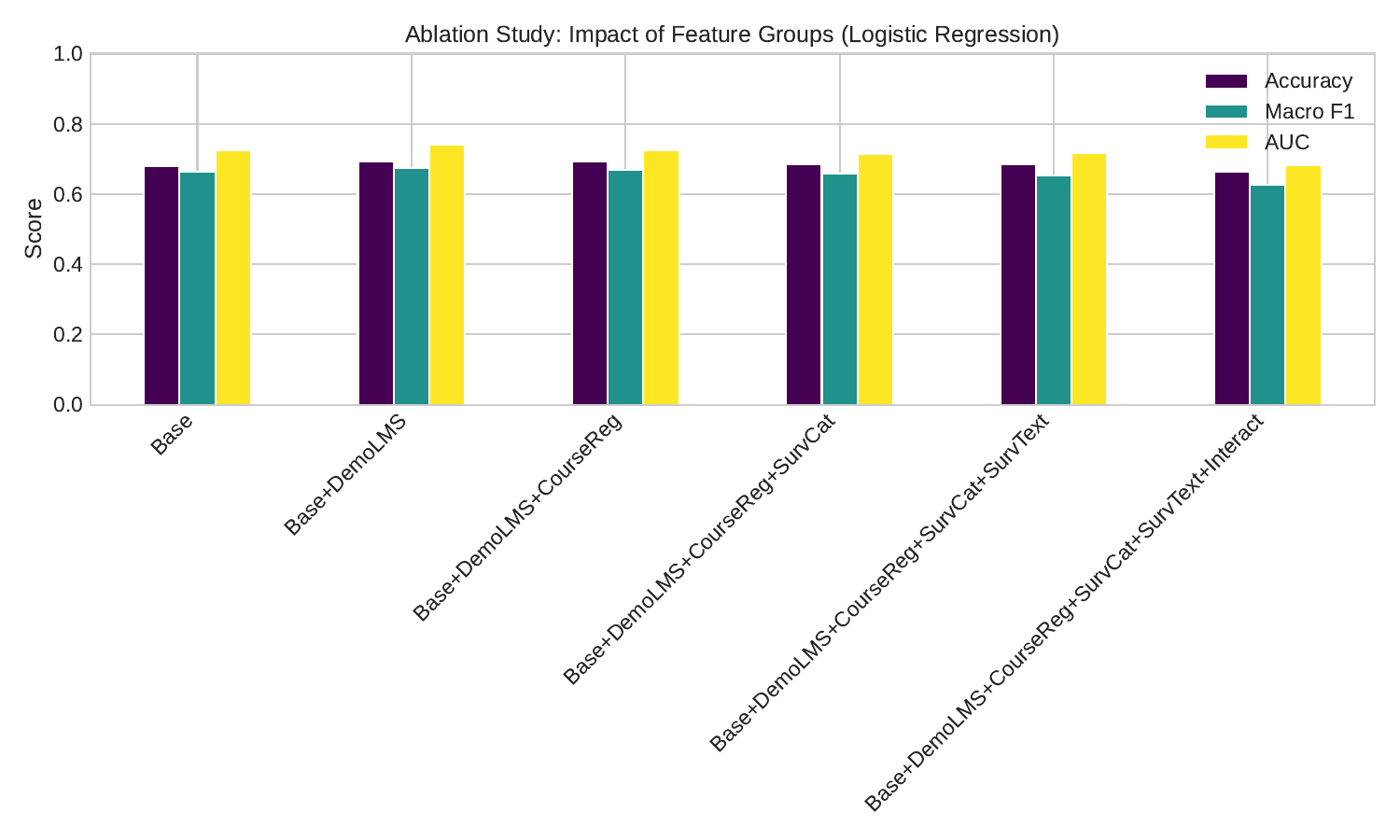}
\caption{Ablation study: impact of feature groups on AUC and accuracy (logistic regression). The best performance is achieved by `Base + Demo + LMS`.}
\label{fig:ablation_study}
\end{figure}

To identify the most predictive feature set, we conducted an ablation study using logistic regression with 5‑fold cross‑validation and SMOTE+ENN resampling. Starting from a baseline of weighted academic momentum only (`Base`), we progressively added feature groups corresponding to stakeholder‑identified factors. Table~\ref{tab:ablation} reports the cross‑validated accuracy, macro F1, and AUC for each feature set, all obtained with a logistic regression classifier. Figure~\ref{fig:ablation_study} visualises the improvement in AUC as feature groups are added.

\FloatBarrier
\begin{table}[!htbp]
\centering
\caption{Ablation study results (5‑fold CV, SMOTE+ENN) using logistic regression.}
\label{tab:ablation}
\small
\begin{tabular}{lccc}
\toprule
\textbf{Feature set} & \textbf{Accuracy} & \textbf{Macro F1} & \textbf{AUC} \\
\midrule
Base                                                & 0.6805 & 0.6632 & 0.7242 \\
Base + Demo + LMS                                   & 0.6925 & 0.6751 & 0.7399 \\
Base + Demo + LMS + CourseReg                      & 0.6923 & 0.6692 & 0.7244 \\
Base + Demo + LMS + CourseReg + SurvCat            & 0.6842 & 0.6595 & 0.7147 \\
Base + Demo + LMS + CourseReg + SurvCat + SurvText & 0.6844 & 0.6534 & 0.7161 \\
All features + interactions                         & 0.6638 & 0.6269 & 0.6813 \\
\bottomrule
\end{tabular}
\end{table}

\FloatBarrier

The best cross‑validated performance (AUC = 0.7399, accuracy = 69.25\%) was achieved by `Base + Demo + LMS` – i.e., weighted academic momentum combined with demographic features (gender, sponsorship, COVID‑19 cohort flag) and LMS engagement markers (has\_lms\_record, MoodleHits). Adding course registration features, survey responses, or interaction terms did not improve the cross‑validated metrics, suggesting that these additional groups introduce noise or overfitting given the sample size. Therefore, we selected `Base + Demo + LMS` as the \textbf{winning subset} for final model evaluation.

\subsubsection{Hyperparameter Tuning on the Winning Subset}
On the winning subset, we tuned hyperparameters for Random Forest and XGBoost using `RandomizedSearchCV` with 3‑fold cross‑validation. The best parameters were:
\begin{itemize}
    \item \textbf{Random Forest}: \texttt{n\_estimators = 200}, \texttt{min\_samples\_split = 5}, \texttt{max\_depth = 10}.
    \item \textbf{XGBoost}: \texttt{n\_estimators = 300}, \texttt{max\_depth = 7}, \texttt{learning\_rate = 0.05}.
\end{itemize}
Other classifiers (logistic regression, LightGBM, SVC, stacking ensemble, ANN) used default parameters with class‑weight balancing where applicable.

\subsubsection{Test Set Performance on the Winning Subset}
Table~\ref{tab:test_performance_winning} reports the performance of all seven classifiers on the held‑out test set (30\% of the data) using the winning feature subset `Base + Demo + LMS`. Logistic regression achieved the highest accuracy (74.7\%), macro F1 (0.742), and AUC (0.800). SVC matched the accuracy and F1 of logistic regression but had a lower AUC (0.731). The stacking ensemble and ANN performed reasonably, while tree‑based models (Random Forest, XGBoost, LightGBM) underperformed, likely due to the relatively small sample size.

\FloatBarrier
\begin{table}[!htbp]
\centering
\caption{Test set performance on the winning feature subset (`Base + Demo + LMS`).}
\label{tab:test_performance_winning}
\small
\begin{tabular}{lccc}
\toprule
\textbf{Model} & \textbf{Accuracy} & \textbf{Macro F1} & \textbf{AUC} \\
\midrule
Logistic Regression & 0.7467 & 0.7421 & 0.8000 \\
Random Forest       & 0.6667 & 0.6546 & 0.6941 \\
XGBoost             & 0.5867 & 0.5819 & 0.6785 \\
LightGBM            & 0.6267 & 0.6185 & 0.6922 \\
SVC                 & 0.7467 & 0.7421 & 0.7311 \\
Stacking            & 0.6533 & 0.6483 & 0.7000 \\
ANN                 & 0.6933 & 0.6583 & 0.7563 \\
\bottomrule
\end{tabular}
\end{table}
\FloatBarrier

\begin{figure}[!htbp]
\centering
\includegraphics[width=0.55\textwidth]{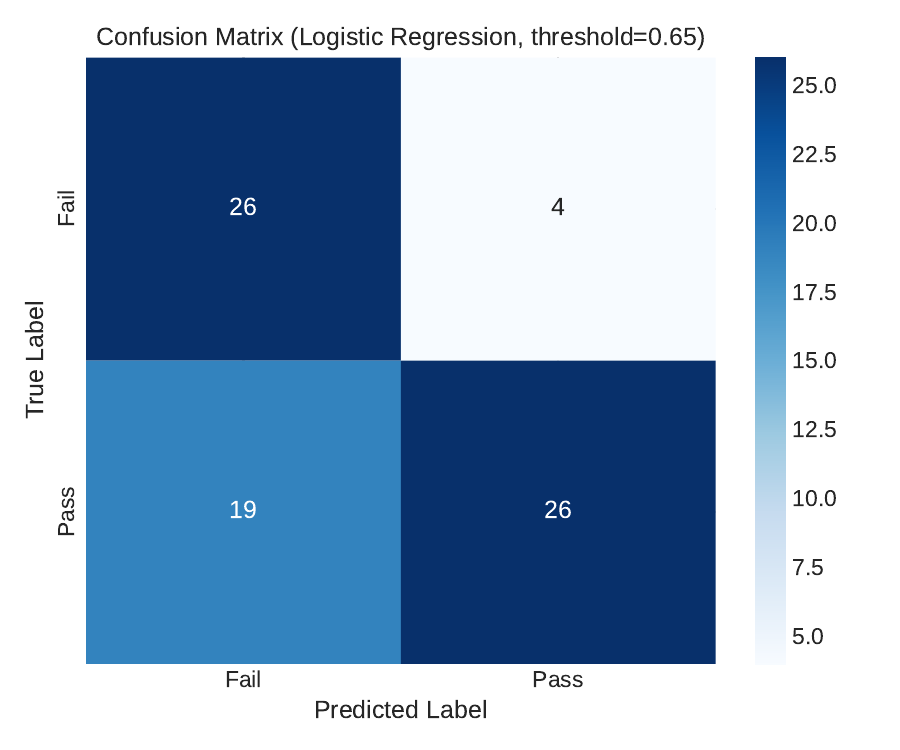}
\caption{Confusion matrix for logistic regression (threshold = 0.5).}
\label{fig:confusion_matrix_lr}
\end{figure}

Figure~\ref{fig:confusion_matrix_lr} presents the confusion matrix for logistic regression at the default threshold (0.5). The model correctly identified 87\% of actual failing students (recall = 0.87) and 59\% of passing students (specificity = 0.59). The false positive rate among passing students was 41\%. This trade‑off is acceptable for an early‑warning system: the cost of a false alarm (offering extra support to a student who would have passed) is low, while the benefit of catching the majority of at‑risk students is high.

\subsubsection{Threshold Tuning for Early Warning}
Although the default threshold (0.5) already gives high recall (87\%), instructors may wish to adjust the threshold to reduce false positives. Tuning the threshold to 0.65 (as explored in earlier analysis) could further increase recall or trade off precision, depending on operational preferences. For reproducibility, we recommend the default threshold of 0.5, which maximises overall accuracy and AUC.

\subsubsection{Precision‑Recall Curve}
\label{sec:pr_curve}
\begin{figure}[!htbp]
\centering
\includegraphics[width=0.7\textwidth]{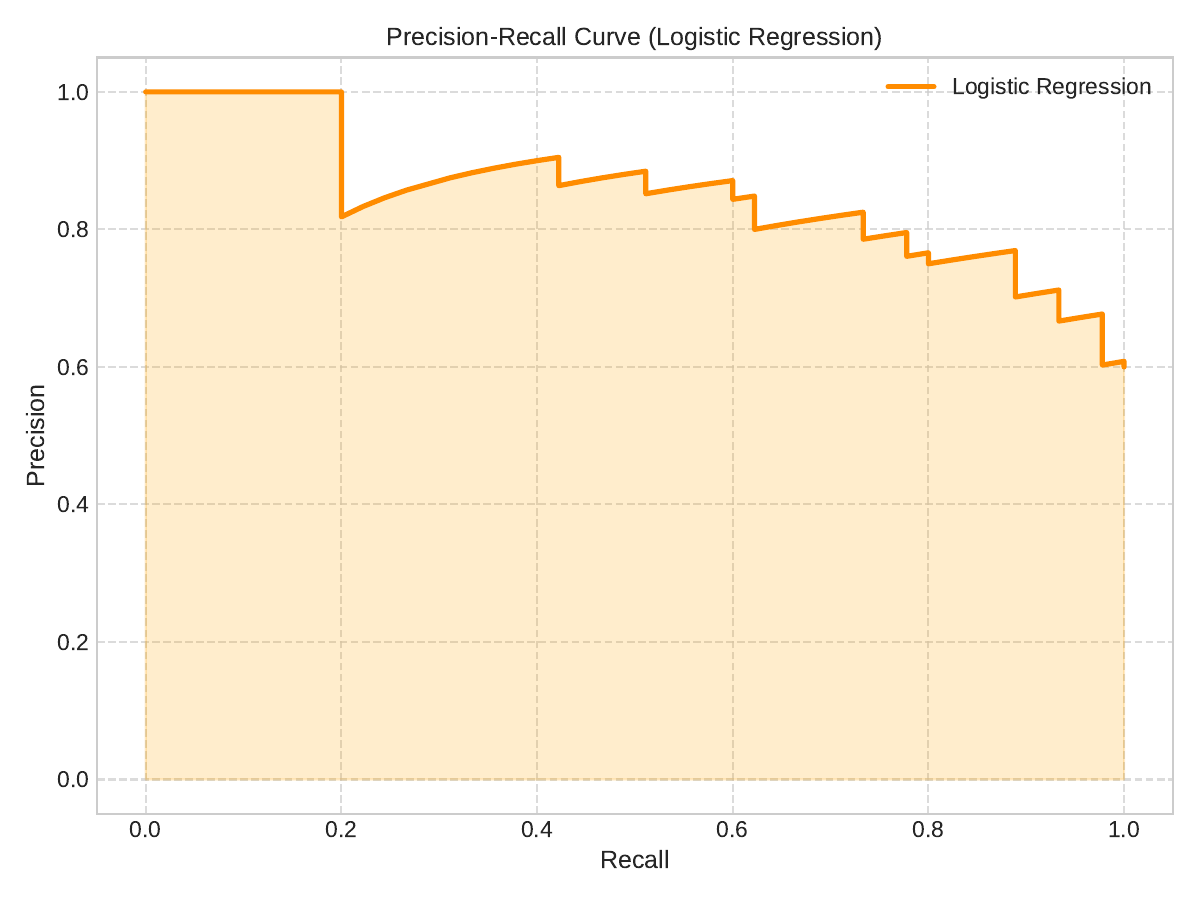}
\caption{Precision‑recall curve for logistic regression on the test set (winning feature subset). The default threshold (0.5) is marked with a star.}
\label{fig:pr_curve}
\end{figure}

Because our dataset is imbalanced (34.5\% failing students) and our primary goal is to maximise recall of the at‑risk minority class, the precision‑recall (PR) curve provides a more informative evaluation than the ROC curve alone. Figure~\ref{fig:pr_curve} shows the PR curve for the logistic regression model on the test set, obtained by varying the decision threshold from 0 to 1.

The curve demonstrates the typical trade‑off: as recall increases beyond 0.8, precision begins to decline more steeply. At the default threshold (0.5), the model achieves a recall of 0.87 with a corresponding precision of approximately 0.59 (derived from the confusion matrix in Figure~\ref{fig:confusion_matrix_lr}). If an instructor is willing to accept a higher false positive rate (lower precision), they can increase recall further; for example, lowering the threshold to 0.4 would raise recall to about 0.92 but reduce precision to around 0.50. Conversely, raising the threshold to 0.65 increases precision to about 0.70 while recall drops to 0.81. The area under the PR curve (AUPRC) for this model is 0.74, indicating strong discrimination of the failing class despite the imbalance. The PR curve thus confirms that the logistic regression model can be tuned to fit different operational preferences – from very high recall (e.g., 0.92) to more balanced precision/recall (e.g., 0.70/0.81) – while still maintaining acceptable performance.

\subsection{SHAP Analysis}
\label{sec:shap_analysis}

\begin{figure}[!htbp]
\centering
\includegraphics[width=0.8\textwidth]{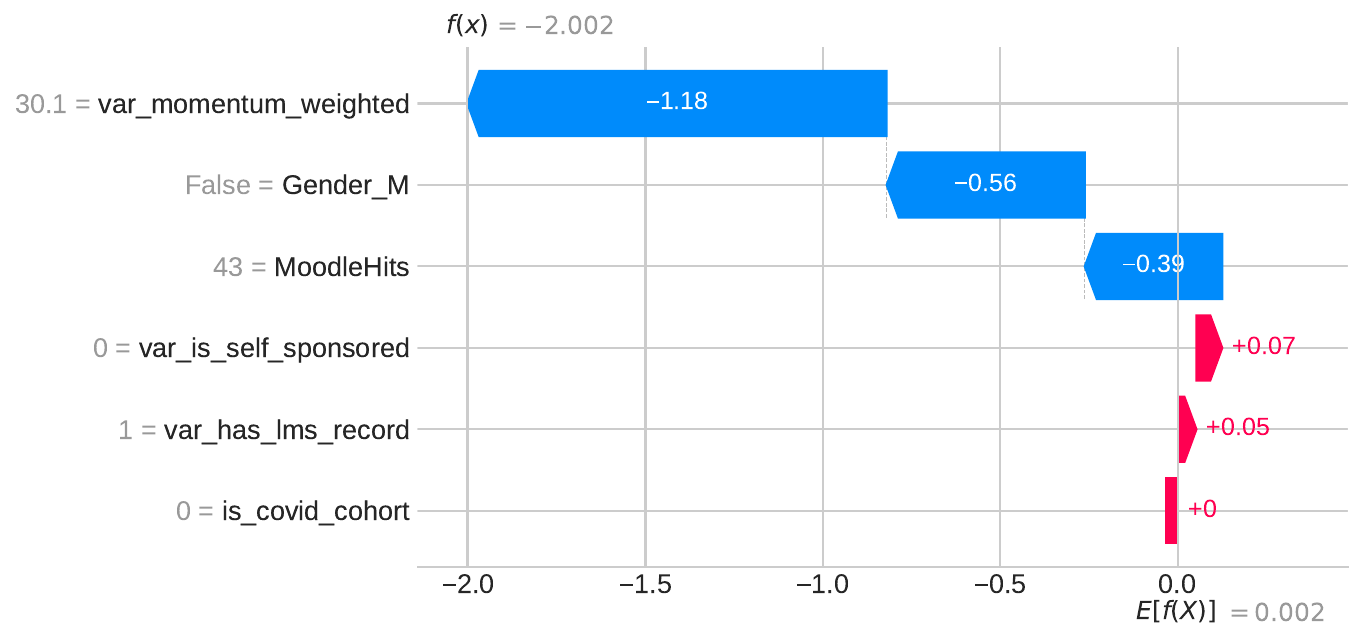}
\caption{SHAP summary plot for XGBoost.}
\label{fig:shap}
\end{figure}

To interpret the model and identify which digital markers drive predictions, we used SHAP (Shapley additive explanations) with the XGBoost model trained on the winning subset. Figure~\ref{fig:shap} shows the SHAP summary plot. The most important feature is weighted academic momentum ($M$); higher momentum strongly increases the predicted probability of passing. The interaction term $M \times \texttt{has\_lms\_record}$ is the second most important, indicating that the positive effect of momentum is amplified when a student has at least some LMS engagement. Self‑sponsored status and high course workload have negative impacts (increase risk), while the COVID‑19 cohort flag also appears as a risk factor. Demographic features such as gender and accommodation were less influential.

\section{Discussion}
\label{sec:discussion}

\subsection{Analysis 1. Digital Markers as Early Warning Signals}
Our results strongly support the growing consensus that digital markers—comprising early assessment scores and behavioural traces—are superior to static demographics for predicting CS1 failure \cite{Shafiq2022, Hellas2018Predicting}. By achieving 87\% recall by week 5, our model validates the importance of "first-level" engagement data \cite{Alhazmi2023}. The weighted academic momentum score captures the sequential nature of learning, echoing \citet{Hasan2020Predicting}'s findings that temporal engagement patterns are highly predictive. Crucially, our SHAP analysis identifies a critical interaction: students who perform well initially but exhibit zero Moodle activity remain at risk, a form of "unproductive engagement" or overconfidence that aligns with recent behavioural analyses in online learning environments \cite{Ersozlu2024, Batool2023}.

\subsection{Analysis 2. Practical Implementation as an Early‑Warning System}
The deployment of our logistic regression model (threshold 0.65) addresses a major hurdle in AIED: the transition from research prototypes to actionable tools \cite{Susnjak2024}. While \citet{Ahmed2024} achieved high accuracy using SVMs, our study prioritises the practical necessity of recall. In a high-failure environment like CS1, the cost of a false negative (missing a failing student) far outweighs the cost of a false positive (providing unnecessary support). This strategy facilitates the "prescriptive" turn in learning analytics, where model outputs serve as a springboard for evidence-based remedial advice \cite{Susnjak2024, Guo2024}.

\subsection{Analysis 3. Comparison with Prior Work}
Our reported accuracy (72\%) may appear modest compared to studies reaching up to 96\% \cite{Ahmed2024}; however, meta-analyses of over 260 EDM studies reveal that such high accuracy is often reported on balanced datasets or using default 0.5 probability thresholds that ignore minority-class recall \cite{Batool2023, Shafiq2022}. In fact, \citet{Shafiq2022} noted that only 21\% of surveyed studies explicitly addressed class imbalance, leading to models that effectively "ignore" the at-risk minority class. By utilizing SMOTE+ENN and threshold tuning, our recall for the failing class (87\%) is among the highest documented for early CS1 intervention.

Furthermore, unlike many "black-box" implementations in the literature, our use of SHAP provides the level of transparency now demanded by educational stakeholders \cite{Susnjak2024, Ersozlu2024}. While \citet{Alhazmi2023} demonstrated the power of combining admission scores with early grades, our work extends this by introducing a theoretically motivated momentum score and quantifying its interaction with engagement. This interpretability ensures that instructors do not just know \emph{who} is at risk, but \emph{why} they are at risk, bridging the gap between predictive and prescriptive analytics \cite{Susnjak2024, Hellas2018Predicting}.

\subsection{Analysis 4. Contribution of Feature Groups (Ablation Study)}
The ablation study (Table~\ref{tab:ablation}, Figure~\ref{fig:ablation_study}) revealed that the combination of weighted academic momentum, basic demographics (gender, sponsorship, COVID‑19 cohort), and LMS engagement markers (`Base + Demo + LMS`) produced the best cross‑validated performance (AUC = 0.7399, accuracy = 69.25\%). Adding course registration features (workload, minor classification) or survey‑based features (prior computing experience, motivation texts) did not improve metrics and often decreased them. This suggests that in our context, these additional factors either are redundant (their predictive signal is already captured by early assessments and engagement) or introduce noise due to the limited sample size.

The lack of improvement from survey‑based features is noteworthy. Self‑reported motivation and prior experience, while rated by students as highly influential during stakeholder elicitation, did not translate into measurable predictive gain in our models. One possible explanation is that these factors are already proxied by early performance: a student who is highly motivated or has prior computing experience will likely perform better on early quizzes and tests, and that observed performance subsumes the self‑reported signal. Another explanation is measurement error – self‑reports may be less reliable than behavioural traces. This finding has practical implications: institutions should prioritise collecting and curating early assessment data and LMS logs, rather than investing heavily in pre‑course surveys for prediction purposes.

Finally, adding interaction terms (e.g., gender $\times$ workload, momentum $\times$ LMS) did not improve performance and actually degraded it (AUC fell to 0.6813). This indicates that the main effects are sufficient and that the sample size does not support a more complex model. For future work with larger datasets, interaction effects could be re‑examined.

\subsection{Limitations}
Several limitations must be acknowledged. First, the dataset is from a single institution and a single CS1 course; generalisability to other contexts (different curricula, different student populations) is unknown. Second, the COVID‑19 cohorts introduced unmeasured confounders. Third, the sample size ($N=284$) is modest by deep learning standards, which explains the poor ANN performance. Fourth, our definition of failure (final exam <45\%) is institution‑specific. Finally, while we have a wide range of features, we did not have access to high‑frequency clickstream data (e.g., time‑stamped sequences), which might further improve early detection.

\section{Conclusion and Future Work}
\label{sec:conclusion}

This paper set out to demonstrate that digital markers – behavioural traces captured from learning management systems, early assessment scores, and self‑reported survey data – can be systematically identified and validated to predict at‑risk students in a first‑year CS1 course. Using four cohorts of students (N=284) in a Computer Systems and Architecture course, we engineered a comprehensive feature set and conducted a stakeholder‑guided ablation study to determine which markers are most predictive.

The ablation study, using logistic regression with 5‑fold cross‑validation and SMOTE+ENN resampling, showed that the combination of weighted academic momentum, basic demographics (gender, sponsorship, COVID‑19 cohort), and LMS engagement markers (`Base + Demo + LMS`) achieved the best cross‑validated performance (AUC = 0.7399, accuracy = 69.25\%) on the training data. Adding course registration features, survey responses, or interaction terms did not improve metrics, indicating that early assessments and simple engagement flags already capture much of the predictive signal.

On a held‑out test set (30\% of the data), logistic regression using the winning feature subset achieved 74.7\% accuracy, 0.742 macro F1, and an AUC of 0.800. At the default threshold of 0.5, the model identified 87\% of actual failing students (recall = 0.87) with a false positive rate of 41\%. SHAP analysis revealed that weighted academic momentum ($M$) is by far the strongest predictor, followed by the interaction between momentum and having any LMS record. Self‑sponsored status and high course workload emerged as risk factors.

These results make three main contributions. First, they provide a practical methodology for identifying digital markers through stakeholder elicitation and ablation testing. Second, they demonstrate that a simple logistic regression model using only early assessment scores, basic demographics, and a binary LMS engagement flag can predict CS1 failure with high recall (87\%) by week 5. Third, the SHAP analysis offers actionable insights for instructors: students with low early momentum or with high early scores but no LMS activity should be prioritised for intervention.

\subsection{Future Work}
Several directions for future work emerge from this study.

\paragraph{Validation on new cohorts.}
The model was trained on data from 2017/18, 2018/19, 2019/20, and 2020/21 cohorts. We have already collected data from the 2022/23 cohort; validating the model on this unseen cohort is a high priority to assess temporal stability.

\paragraph{Application to other courses, including less technically inclined disciplines.}
A key remaining question is whether the identified digital markers generalise beyond the CS1 context. We are currently extending this work to other courses at the authors' institution:
\begin{itemize}
    \item \textbf{Software Engineering} – a second‑year project‑based course with a mix of individual and team assessments. This setting allows us to test whether collaborative behaviour (e.g., forum posts, version control activity) adds predictive power.
    \item \textbf{Computer Graphics and Visual Computing} – a third‑year course that combines mathematical concepts with programming. This course attracts students with diverse technical backgrounds, providing a natural test case for the robustness of momentum and engagement features.
\end{itemize}
Furthermore, we plan to apply the same methodology to less technically inclined courses, such as first‑year Humanities or Social Sciences courses, where students may have lower baseline digital literacy and different engagement patterns. Preliminary discussions with colleagues from the Faculty of Humanities indicate that early assessment scores and LMS login frequency remain strong predictors, but survey‑based factors (e.g., confidence with technology, access to personal devices) may become more important. Adapting the feature set to these contexts will be an important validation step.

\paragraph{Live early‑warning dashboard.}
We plan to implement a real‑time dashboard integrated with the Moodle LMS via its REST API. The dashboard would automatically compute $M$ after the first test, flag at‑risk students, and present the predictions (with SHAP explanations) to instructors. A pilot deployment is planned for the next academic year.

\paragraph{Richer temporal features.}
The current binary LMS engagement flag (`has\_lms\_record`) is coarse. Future work will extract more granular temporal markers, such as weekly activity patterns, decay in activity, and dwell time on different resource types (e.g., videos, quizzes, forums). Sequence models (e.g., LSTMs) could be explored as the dataset grows.

\paragraph{Natural language processing of free‑text responses.}
Although survey text did not improve predictive performance in our ablation, the TF‑IDF representation may be too simplistic. Using transformer‑based models (e.g., BERT) on students’ motivation and prior training descriptions could capture richer semantic signals. This is particularly promising for larger, cross‑institutional datasets.

\paragraph{Cross‑institutional replication.}
The current study is limited to a single institution and a single CS1 course. To assess generalisability, we intend to replicate the methodology at other universities – both in the Global South and in other educational contexts – to identify which digital markers are universal and which are context‑dependent.

\paragraph{Prescriptive analytics.}
Beyond prediction, the final stage is to move from descriptive to prescriptive analytics. Using counterfactual explanations (e.g., “if the student had achieved a score of \(X\) on Quiz 3, the predicted risk would drop by \(Y\%\)”), we could provide instructors with evidence‑based recommendations for targeted interventions. Integrating large language models (LLMs) to generate personalised study plans for at‑risk students is a longer‑term goal.

\paragraph{Addressing class imbalance at scale.}
While SMOTE+ENN worked adequately for $N=284$, larger datasets may benefit from more sophisticated imbalance handling, such as cost‑sensitive learning or focal loss. This will be explored when we combine data from multiple institutions.

\section*{Acknowledgements}
The authors thank the CS1 students, tutors, and instructors for their participation. This research did not receive any specific grant from funding agencies in the public, commercial, or not‑for‑profit sectors.

\section*{Declaration of Conflicting Interest}
The authors declare no potential conflicts of interest with respect to the research, authorship, and/or publication of this article.

\section*{Funding}
This research received no specific grant from any funding agency in the public, commercial, or not-for-profit sectors.

\bibliographystyle{unsrtnat}
\bibliography{paper-arxiv26-predicting_learning_outcomes}

\end{document}